# Controlling X-ray emission with dispersion-engineered surface plasmon polaritons

H. Aknin[1], Y. Klein[1,2], S. Shwartz [1*]

[1]*Physics Department and Institute of Nanotechnology, Bar-Ilan University, Ramat Gan, 52900 Israel*

[2]*Department of Physics, University of Ottawa, Ottawa, Ontario K1N 6N5, Canada*

We propose controlling the angular and spectral distribution of hard x-ray emission by entangling x-ray photons with ultraviolet surface plasmon polaritons (SPPs) whose dispersion is engineered by a metal-dielectric multilayer on a nonlinear crystal. Spontaneous parametric down-conversion of an x-ray pump produces a hard x-ray signal photon correlated with an ultraviolet SPP mode near its resonance. Engineering the SPPs dispersion reshapes the phase-matching landscape and imprints tunable angular-spectral structure on the emitted x-ray photons. The scheme enables compact, designable control of x-ray emission and extends surface-plasmon-assisted nonlinear and quantum x-ray optics.

Controlling the propagation direction and spectrum of electromagnetic radiation is central to modern science and technology, underpinning applications ranging from precision sensing to information processing and communication [1]. At optical frequencies, such control is routinely achieved by engineering light-matter interactions mediated by valence electrons, for example, using dispersion engineering, nonlinear optical processes, and plasmonic confinement [2]. In the hard x-ray regime, however, the large photon energy and momentum render the coupling to valence electrons intrinsically weak, leaving most materials with refractive indices close to unity and severely limiting the degrees of freedom available for shaping x-ray radiation.

Recently, we experimentally demonstrated control over the emission direction and spectrum of hard x-rays by entangling x-ray photons with ultraviolet surface plasmon polaritons (SPPs) [3]. In that work, the entanglement was generated via spontaneous parametric down-conversion (SPDC) of an x-ray pump into an x-ray signal photon and a near-field ultraviolet SPP at the surface of an aluminum crystal. Because the signal is born entangled with the SPP, energy and transverse-momentum conservation map the SPP dispersion onto the angular-spectral distribution of the emitted x-rays. The accessible tuning range in that proof-of-principle experiment, however, was largely set by the intrinsic material response, leaving only limited control via the choice of crystal or the dielectric environment.

Here we extend this concept by introducing a dispersion-engineered metal-dielectric multilayer stack that tailors the ultraviolet SPP dispersion. This added design freedom enables deterministic and reconfigurable control over the x-ray emission directionality and spectrum, without modifying the underlying nonlinear crystal.

We note that a recent theoretical study proposed a related route to controlling x-ray emission by exploiting x ray to optical SPDC in nonlinear photonic crystals, where photonic dispersion is engineered to shape the generated radiation [4]. In contrast, we shape the x-ray emission by engineering the dispersion of a surface-bound plasmonic partner, rather than a propagating optical photon, enabling control through near-field confinement at a metal interface. More broadly, several experiments over the past decade have demonstrated nonlinear wave-mixing processes that couple hard x-rays to longer-wavelength fields [5–12]. While the conversion efficiencies in this regime remain low, they are nevertheless measurable and sufficient to reveal rich physics.

We consider an x-ray emission control scheme as is illustrated in Fig. 1. The structure comprises a stack of thin non-crystalline aluminum (Al)-alumina ($Al_2O_3$) bilayers on the top of an Al crystal. This structure is illuminated by an x-ray pump beam at frequency $\omega_p$ and incident angle $\theta_p$, generating signal and idler photons at $\omega_s$ and $\omega_i$ through nonlinear interaction in the crystal. The signal photon is emitted close to the Bragg direction and is parameterized by small angular deviations: an in-scattering-plane deviation $\delta\theta_s$ relative to the Bragg angle, and an out-of-scattering-plane deviation $\delta\phi_s$, relative to the x-z plane.

When the frequency of the idler photon matches the surface plasmon resonance (SPR) frequency and its in-plane momentum matches that of the SPP mode, strong coupling occurs, and the idler forms an SPP [3]. Through conservation laws, the multilayer controlled SPP dispersion is directly imprinted onto the angular distribution and spectrum of the generated entangled x-ray signal photon [3]. The x-ray emission is therefore dictated by the generalized SPP dispersion relation of the multilayer.

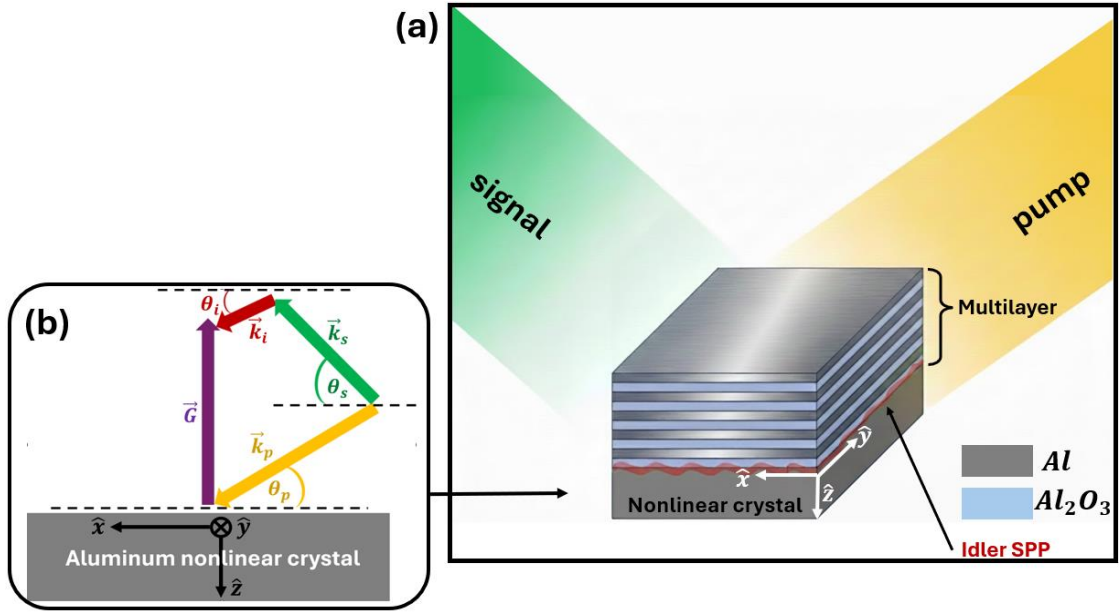

**Fig. 1: (a)** schematic of x-ray-to-SPP SPDC with dispersion engineering. A monochromatic x-ray pump beam impinges on an aluminium nonlinear crystal at incidence angle $\theta_p$, generating an x-ray signal photon and an idler excitation that couples to a dispersion-engineered SPP at the metal-dielectric interface. **(b)** Phase-matching diagram, showing the k-vectors of the pump, signal and idler, k$_p$, k$_s$, k$_i$ and the reciprocal lattice vector G (purple). Angles are defined with respect to the atomic planes (dashed lines).

To model SPDC in a lossy dispersive medium, we use macroscopic quantum electrodynamics (MQED), which quantizes the macroscopic Maxwell equations for the electric-field operators in absorbing media [13,14]. Dissipation is incorporated via a Langevin formulation by introducing noise current operators, $\hat{j}_N$. Together with the nonlinear current operators, $\hat{j}_{NL}$, the total source current driving the field at the $u^{th}$ mode is given by $\hat{j}_u = \hat{j}_{NL,u} + \hat{j}_{N,u}$, where $u = s, i$ denote the signal and idler modes, respectively [3,13,14].

The signal and idler electric-field operators, $\hat{E}_u(\vec{r}, \omega_u)$, then satisfy:

$$\nabla \times \nabla \times \hat{E}_u(z, \omega_u) - \frac{\omega_u^2}{c^2}\varepsilon(z, \omega_u)\hat{E}_u(\vec{r}, \omega_u) = -i\omega_u\mu_0\hat{j}_u(\vec{r}, \omega_u), \quad (1)$$

where $\varepsilon(z, \omega_u)$ is the position dependent permittivity.

To solve Eq. (1) for the electric-field operators, we note that the penetration depth of the idler in the Al nonlinear crystal is much shorter than its wavelength. Consequently, at the idler frequency the nonlinear current density is much smaller than the Langevin noise current, $\hat{j}_{NLi} \ll \hat{j}_{Ni}$ [3]. In this limit, the idler field is dominated by vacuum fluctuations, and its correlation function is determined by the imaginary part of the electromagnetic Green's tensor, $G(r, r', \omega)$, defined by the Maxwell equation for $\hat{E}_i$ [3].

This approximation decouples the signal and idler equations, since the nonlinear mixing contribution to the idler field is negligible. We therefore solve first for $\hat{E}_i$ and then substitute it into the signal-field equation to obtain $\hat{E}_s$.

The x-ray signal count rate, $\Gamma_s$, is calculated using the standard photon number operator in the Heisenberg picture, $\Gamma_s = \langle 0|\hat{a}_s^\dagger \hat{a}_s|0\rangle$, where $\hat{a}_s^\dagger$ and $\hat{a}_s$ are the creation and annihilation operators for the signal mode at the output of the crystal and are related to the signal electric field operators $\hat{E}_s$ by: $\hat{E}_s(q_s, \omega_s, z) = \sqrt{\frac{2\hbar\omega_s\eta(\omega_s)}{\sin(\theta_s)}}\hat{a}_s(q_s, \omega_s, z)$, where $\eta(\omega_s)$ is the impedance of the signal field [3].

The count rate of the x-ray signal photons is given by [3]:

$$\Gamma_s = \int d\omega_i \int \frac{d^2q_s}{(2\pi)^2} \frac{\hbar\omega_i^2}{\pi\epsilon_0 c^2} |C(q_s, q_p, \omega_i)|^2 \frac{q_i^2}{k_i^2(\omega_i)} \times \\ \int_L^0 dz' \int_L^0 dz\, e^{i\Delta k_{p,s}(z-z')} \text{Im}\{g_{q_i}(z, z', \omega_i)\}, \quad (2)$$

where $L$ is the thickness of the nonlinear crystal. The integration ranges of the idler frequency $\omega_i$ and the signal transverse wave vector $\vec{q}_s$ are set by the energy bandwidth and the angular acceptance of the detector. Transverse momentum conservation along the surface of the nonlinear crystal imposes $\vec{q}_p = \vec{q}_s + \vec{q}_i$.

The scalar Green's function $g_{q_i}$ is related to $G(r, r', \omega)$ by a Fourier transform over the transverse momentum and projection onto the idler polarization: $G(r, r', \omega_i) = \int \frac{d^2\vec{q}_i}{(2\pi)^2} e^{i\vec{q}_i\cdot(\vec{\rho}-\vec{\rho}')} g_{q_i}(z, z', \omega_i)$. For a planar interface at $z = 0$, it contains a bulk term and a reflected term determined by the multilayer reflection coefficient [3]. Consequently, the properties of the bilayer structure and its materials enter the signal count rate through $g_{q_i}$ [3]. The longitudinal phase mismatch is $\Delta k_{p,s} = k_{p,z}(q_p, \omega_{p,0}) + k_{s,z}(q_s, \omega_{p,0} - \omega_i) - G$, where $k_{p,z}$ and $k_{s,z}$ are the longitudinal pump and signal wave vectors, respectively, and $G$ is the reciprocal lattice vector used for phase matching. The idler contribution is included implicitly via $g_{q_i}$. The prefactor $C(q_s, q_p, \omega_i)$ accounts for the propagation angles of the pump and signal photons and includes the nonlinear coupling coefficient of the nonlinear current at the signal mode $\hat{J}_{NL,s}$.

Next, we consider the following parameters in our simulation: we assumed the nonlinear crystal is illuminated with a pump beam of energy $\hbar\omega_p = 10\,KeV$ and a flux of $2 \times 10^{12} \left[\frac{photons}{sec}\right]$ and that phase matching is achieved using the reciprocal-lattice vector normal to the Al (0,2,2) planes.

We first illustrate how our scheme with these parameters controls the emission of the signal x-ray field. Fig. 2(a) shows the signal intensity as a function of detector angle and idler energy. A prominent feature is the band-like dispersion of the x-ray signal photons. **This structure arises from the plasmonic multilayer, which reshapes the joint signal-idler phase-matching landscape.**

In Fig. 2(b) we show the angular distribution of the x-ray signal photons in the $(\delta\theta_s, \delta\phi_s)$ plane, evaluated for signal photons selected along the green dashed trajectory in Fig. 2(a) and centred at $\hbar\omega_s = \hbar\omega_p - 6$ eV (corresponding to an idler energy of 6 eV). Of importance, this angular distribution is energy dependent. **By selecting a different signal energy (corresponding to shifting the green dashed line in Fig. 2(a) to up or down) we can modify the angular distribution of the emitted x-ray signal**.

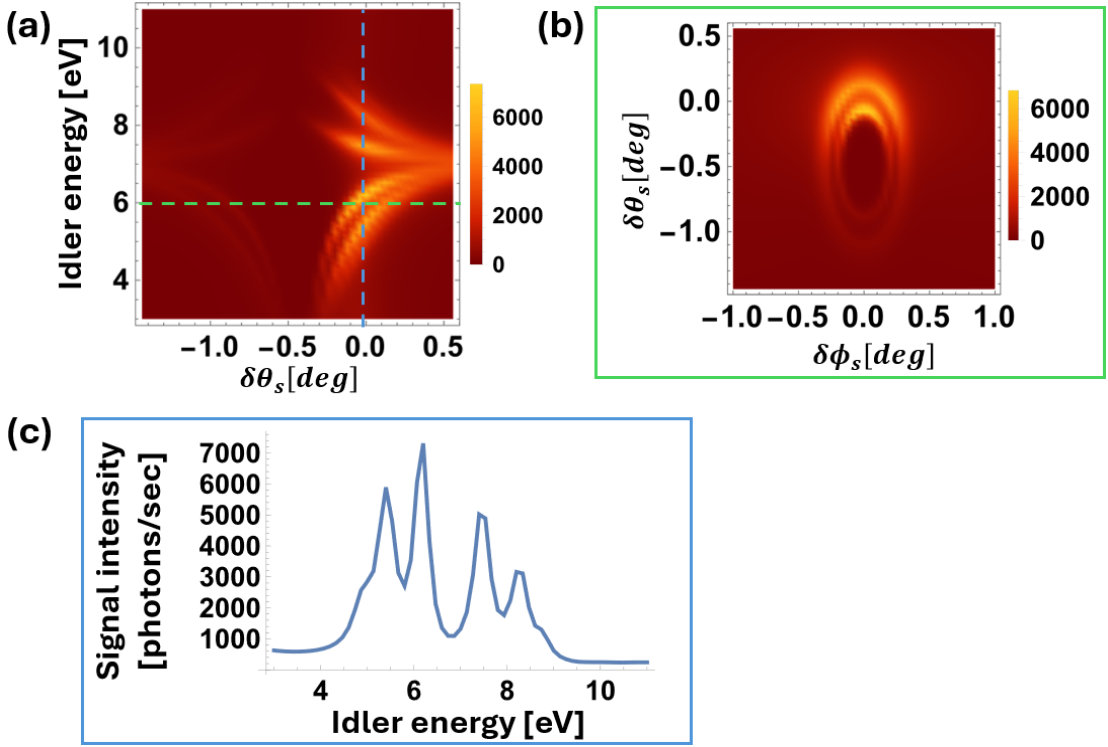

**Fig. 2:** (a) Simulated signal intensity as a function of the detector angle and idler photon energy for a metal-dielectric multilayer comprising three Al layers separated by Al$_2$O$_3$, with an identical thickness of $d_{Al} = d_{Al_2O_3} = 7\ nm$. (b) Signal angular spectrum for $\hbar\omega_i$=6 eV, corresponding to the green dashed line in (a). (c) Signal energy spectrum for $\delta\theta_s = -0.02°$ corresponding to the blue dashed line in (a).

Fig. 2(c) presents the corresponding spectral shaping. We plot the signal intensity as a function of idler energy $\hbar\omega_i$ for signal photons emitted along the blue dashed trajectory in Fig. 2(a), centred at $\delta\theta_s = -0.02°$ and $\delta\phi_s = 0$. **As is evident from Fig. 2(a), selecting different angular trajectory (corresponding to shifting the blue dashed line in Fig. 2(a) to the left or to the right) leads to a different emission spectrum**.

In addition, distinct regions of strongly suppressed emission are evident across the angle-energy maps. These correspond to plasmonic bandgaps, i.e., parameter ranges in which no allowed idler SPP mode exists in the multilayer, thereby "blocking" SPDC emission into the corresponding signal angles and energies. In this way, the multilayer functions as a combined angular and spectral filter with near-unity extinction in selected regions of phase space.

Additional control is achieved by varying the number of bilayers. Fig. 3 presents the emission angles and idler energies for several bilayer thicknesses. It is prominent that varying the number of bilayers primarily controls the modal multiplicity. As $N_{bilayers}$ increases from 1 to 3 to 5, additional rings appear in the signal angular spectra [Figs. 3(a–i)], reflecting the emergence of additional dispersion branches. Consistently, additional resonances appear in the energy spectra. Increasing $d_{Al_2O_3}$ reduces the characteristic scattering-angle radius in Figs. 3(a–i). Spectrally, the low-energy resonances blue-shift and weaken, while higher-energy resonances red-shift and strengthen as they approach $\hbar\omega_i = 7$ eV, the SPR of a single Al/Al$_2$O interface. The net effect is a tighter clustering of both angular and spectral features.

This trend can be understood by considering the limit $d_{Al_2O_3} \to \infty$, where the scheme effectively reduces to a single Al/Al$_2$O interface and supports a single SPP branch. That branch exhibits the smallest in-plane scattering-angle radius over the full idler-energy range and an SPR energy near 7 eV. Accordingly, as $d_{Al_2O_3}$ increases, the multilayer branches converge toward the single-interface SPP dispersion.

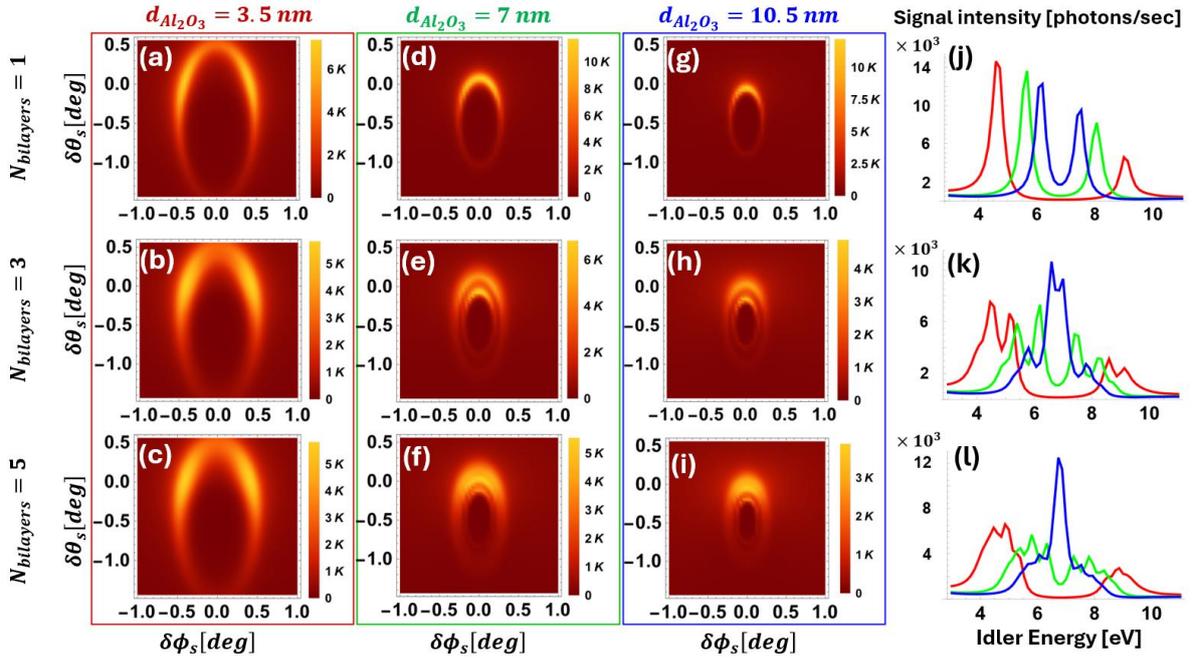

**Fig.3:** Signal angular spectrum at $\hbar\omega_i = 6$ eV for three $d_{Al_2O_3}$ thicknesses: **(a)-(c):** $d_{Al_2O_3} = 3.5\ nm$, **(d)-(f):** $d_{Al_2O_3} = 7\ nm$ and **(g)-(i):** $d_{Al_2O_3} = 10.5\ nm$. Signal energy spectrum for $\delta\theta_s = -0.02°$ and $\delta\phi_s = 0$ is shown in **(j)-(l)**. In all panels the number of bilayers is 1 (top row), 3 (middle row), and 5 (bottom row).

Finally, Fig. 3 demonstrates that the multilayer can be engineered to enhance or suppress the signal. For example, with five bilayers the signal count rate at 7 eV increases by a factor of 127 by changing the dielectric layers thickness from 3.5 nm to 10.5 nm.

In conclusion, we show that hard-x-ray emission can be controlled by coupling SPDC to dispersion-engineered ultraviolet SPPs. By tailoring the multilayer-supported SPP dispersion, the phase-matching landscape is reshaped and the tunable angular-spectral structure is imprinted on the emitted x-ray signal. Beyond providing a compact route to shaping x-ray emission beyond conventional x-ray optics, the approach enables the design of single-photon x-ray sources with tailored directional and spectral distributions, including lens-like concentration of emission into selected angular modes.

Here we have illustrated the concept in aluminium, but the approach can be extended to other plasmonic platforms to target different idler-frequency ranges. Looking ahead, dispersion engineering of the correlated long-wavelength partner offers new degrees of freedom for enhancing x-ray/low-frequency nonlinear coupling and for generating coherent or quantum x-ray states with structured spatiotemporal and spectral wave functions. Finally, because the penetrating x-ray photons remain correlated with a near-field SPP that is highly sensitive to local boundary conditions, the mechanism offers a pathway to probing the optical response of thin or buried layers embedded in otherwise optically opaque materials.